\documentclass[10pt,letterpaper]{article}

\makeatletter
\def\input@path{{style/}}
\makeatother

\usepackage[T1]{fontenc}
\usepackage{times}
\usepackage{url}
\usepackage{graphicx}
\usepackage{caption}
\usepackage{subfig}
\usepackage{xspace}
\usepackage{color}
\usepackage{setspace}
\usepackage{balance}

\usepackage{amsmath, amssymb, amsthm, bm, bbm}
\usepackage{style/mysymbol}
\usepackage{psfrag}
\usepackage{array}

\allowdisplaybreaks[1]

\usepackage[linesnumbered,ruled,vlined]{algorithm2e}
\usepackage{algpseudocode}

\usepackage{booktabs, tabularx, multirow, tabu, array}
\usepackage{diagbox}
\usepackage[table,xcdraw]{xcolor}
\usepackage{slashbox}

\usepackage{indentfirst}
\usepackage{adjustbox}

\usepackage{verbatim}
\usepackage{fancybox}
\usepackage{secdot}     
\usepackage{style/acl2015}
\usepackage{footnote}
\usepackage[colorinlistoftodos]{todonotes}

\newtheorem{remark}{\bfseries Remark}
\input{mysymbol.sty}

\DeclareRobustCommand{\edit}[1]{\textcolor{black}{#1}}

\def\BibTeX{{\rm B\kern-.05em{\sc i\kern-.025em b}\kern-.08em
    T\kern-.1667em\lower.7ex\hbox{E}\kern-.125emX}}

\begin{document}
    
\title{
    \texttt{NEO-Grid}:
A \underline{Ne}ural Approximation Framework for \\ \underline{O}ptimization and Control in Distribution \underline{Grid}s
}

\author{ \normalfont Mohamad Chehade and Hao Zhu \\
Chandra Family Department of Electrical and Computer Engineering\\
The University of Texas at Austin \\
{{\{chehade, haozhu\}@utexas.edu}} \\
}


 \maketitle


\begin{abstract}
The rise of distributed energy resources (DERs) is reshaping modern distribution grids, introducing new challenges in attaining voltage stability under dynamic and decentralized operating conditions. This paper presents \texttt{NEO-Grid}, a unified learning-based framework for volt-var optimization (VVO) and volt-var control (VVC) that leverages neural network surrogates for power flow and deep equilibrium models (DEQs) for closed-loop control. Our method replaces traditional linear approximations with piecewise-linear ReLU networks trained to capture the nonlinear relationship between power injections and voltage magnitudes. For control, we model the recursive interaction between voltage and inverter response using DEQs, allowing direct fixed-point computation and efficient training via implicit differentiation. We evaluated NEO-Grid on the IEEE 33-bus system, demonstrating that it significantly improves voltage regulation performance compared to standard linear and heuristic baselines in both optimization and control settings. Our results establish NEO-Grid as a scalable, accurate, and interpretable solution for learning-based voltage regulation in distribution grids.
\end{abstract}

\setcounter{section}{0}  

\section{Introduction}
\label{sec:intro}

\textcolor{black}{
Distribution grids are undergoing a rapid transformation driven by the proliferation of distributed energy resources (DERs) such as solar photovoltaics, electric vehicles, and battery storage. This evolution introduces new operational challenges for voltage regulation, due to the increased variability and heterogeneity of DERs. Meanwhile, the rise of DERs opens up new opportunities for \emph{volt–var management}, thanks to their ability to fast vary the reactive power outputs by controlling the power-electronic interface \cite{IEEE1547,NREL1547Overview2021}.}

\textcolor{black}{
There are two principal modes of volt–var management. In \textbf{volt–var optimization (VVO)}, reactive power injections are computed using a centralized optimization to minimize system-wide voltage deviation \cite{Jahangiri2013,DrZhu}\textcolor{black}{\cite{Jha2019_BiLevelVVO,Savasci2022_TIA_VVO,Ahmadi2023_RobustVVO,Long2022_BiLevelVVO}}. In contrast, the \textbf{volt–var control (VVC)} seeks local, autonomous policies that map bus-level voltage measurements to reactive power decisions \cite{DrZhu,Jahangiri2013}. The stability and optimality performances of local droop-based designs  can be analyzed; see e.g.,  \cite{FarivarLow2013,DrZhu}. Both approaches hinge on an accurate modeling of voltage responses to power injections, namely, the power-flow (PF) model.}

\textcolor{black}{
Because the PF models for AC circuits are governed by nonlinear equations, linearized approximations have been commonly advocated such as the well-known LinDistFlow model \cite{distflow,IEEE33BW} and its three-phase counterpart \cite{aa}. While such models enable tractable optimization, they may overlook important nonlinearities, especially under heavy loading or in highly resistive feeders.}
To bridge this gap, we adopt a unified, data-driven design that builds a \emph{neural PF surrogate}, termed the \texttt{NEO-Grid} model. It provides an accurate yet tractable ``neuro-simulator'' which is applicable to tackle a variety of distribution decision-making and predictive-analysis tasks. Specifically, ReLU-based neural networks (NNs) have proven effective in approximating PF mappings \cite{Donon2019,PowerFlowNet2024,Young-ho} and producing tractable optimization via mixed-integer linear (MIL) reformulations for the resultant piecewise-linear (PWL) activations \cite{PanDuriez2020}. This surrogate-centric view enables to optimize  both the var decision variables (VVO) and control rules (VVC) with the \emph{same} learned model. Furthermore, this \texttt{NEO-Grid} model could be extended to perform predictive modeling, sensitivity analysis, what-if studies, as well as broader digital-twin use cases; see an overview in \cite{LiaoGNNReview2021}.

\textcolor{black}{
Nonetheless, designing closed-loop VVC rules for \emph{optimal equilibrium} performance  remains computationally challenging due to the recursive interactions between voltages and var controls: per the VVC rule, voltages drive the var outputs, which in turn reshape feeder-wide voltages following the PF model. \edit{color=black}{Prior solution uses unrolled recurrent NNs (RNNs) which are known for the memory burden and gradient vanishing issue.} We instead leverage the implicit learning toolbox and \textbf{Deep Equilibrium Model (DEQ)}  \cite{DEQ,MDEQ}, which can efficiently optimize the fixed-point equilibrium for the composed layers of both neural PF mapping and VVC rules. By using root-finders such as Anderson acceleration \cite{Anderson1965,WalkerNi2011}, DEQs utilizes the \emph{implicit differentiation} at the equilibrium for scalable training on the single-layer policies, without the need of using the long trajectories in RNNs.}

\textcolor{black}{
Our contributions are threefold:~i) \emph{Unified PF design}—we construct a single neural PF surrogate that underpins the need for accurate yet tractable solutions for both paradigms of distribution optimization (VVO) and control (VVC); ii) \emph{Generalizability}—our ``universal'' neuro-simulator, \texttt{NEO-Grid}, is promising for supporting diverse distribution tasks including data-fusion and predictive analysis within the digital-twin workflow; and iii) \emph{Equilibrium-based VVC learning}—we combine the PWL VVC rule representation with the \texttt{NEO-Grid} based PF surrogate to allow for an efficient optimization of the closed-loop control performance via DEQ-based implicit differentiation. This proposed VVC design outperforms existing linear/RNN-based methods \cite{FarivarLow2013,Kekatos1,Kekatos2} and approaches the optimal performance for the actual nonlinear system.}

\hfill 

\section{System Modeling}
\label{sec:system_modeling}


We consider a distribution system consisting of a substation bus feeding  $N$ buses in the set $\mathcal{N} = \{1,\ldots,N\}$. We assume a radial, single-phase system for simplicity, and thus the number of lines is also $N$. However, the proposed ML and decision-making techniques are not restricted to this setup and can be generalized to non-radial, multi-phase networks. A subset of buses in $\mathcal{D} \subset \mathcal{N}$ hosts  distributed energy resources (DERs), such as inverter-interfaced generation or battery. Each node $i \in \mathcal{N}$ is connected to the active and reactive power demand denoted by $(p_i^c, q_i^c)$ and generation by $(p_i^g, q_i^g)$. In addition, let $v_i$ denote the  voltage magnitude per bus $i$, and assume that the substation voltage $v_0$ is fixed and known. 
 We can stack all voltage/power variables into $N\times 1$ vectors, namely, $\bbv$, $\bbp$, and $\bbq$. Hence, the real power injection $\bbp := \bbp^g-\bbp^c$; and similarly for $\bbq :=\bbq^g-\bbq^c$. 

Let us use $\bbv = f(\bbp, \bbq)$ to represent the  nonlinear AC-PF model. To form $f(\cdot)$, we use the the network topology given by the $N$ line segments collected in the set $\mathcal{E} \subset \mathcal{N} \times \mathcal{N}$, as well as the resistance $r_{ij}$ and reactance $x_{ij}$ for each line $(i,j) \in \mathcal{E}$. Let $p_{ij}$ and $q_{ij}$ respectively denote the real and reactive power flowing from bus $i$ to bus $j$ on line $(i,j)$. This allows one to express the nonlinear PF model $f(\cdot)$ using the well-known DistFlow equations \cite{distflow}:
\begin{subequations}
    \label{eq:DF}
\begin{align}
    p_{ij} &= \sum_{k \in \mathcal{N}_j} p_{jk} + r_{ij} \textstyle \frac{p_{ij}^2 + q_{ij}^2}{v_i^2} + p_j^g - p_j^c, \\
    q_{ij} &= \sum_{k \in \mathcal{N}_j} q_{jk} + x_{ij} \textstyle \frac{p_{ij}^2 + q_{ij}^2}{v_i^2} + q_j^g - q_j^c, \\
    v_j^2 &= v_i^2 - 2(r_{ij} p_{ij} + x_{ij} q_{ij}) + (r_{ij}^2 + x_{ij}^2) \textstyle \frac{p_{ij}^2 + q_{ij}^2}{v_i^2}.
\end{align}
\end{subequations}
Here, $\mathcal{N}_j$ denotes the set of downstream child nodes of bus $j$. The fractional term $(p_{ij}^2 + q_{ij}^2)/v_i^2$ represents the squared line current magnitude, corresponding to the power losses on line $(i,j)$. 
To simplify \eqref{eq:DF}, the linearized Distflow (LinDistFlow) model is widely used, which ignores all fractional terms \cite{IEEE33BW}. This will be discussed more in Section \ref{sec:learning_pf} later on.



Optimizing feeder voltage using the flexibility of  $\bbq^g$ from smart inverters along with other reactive power (var) resources is becoming increasingly important to mitigate the volatility of distribution systems. We consider a basic version of this \textbf{volt-var optimization (VVO)} problem \cite{Yize,Kekatos1,DrZhu}, that seeks to minimize the voltage deviation from the nominal 1.0 pu, given by  
\begin{subequations}\label{eq:vvo}
\begin{align}
    \min_{\bbq^g} \quad & \| \bbv - \mathbf{1} \|_2^2 \label{eq:vvo_o} \\
    \text{s.t.} \quad & \bbv = f(\bbp^g - \bbp^c, \bbq^g - \bbq^c) \\
                      & \underline{\bbq} \leq \bbq^g \leq \bar{\bbq}.
\end{align}
\end{subequations}
Here, the range $[\underline{\bbq}, \bar{\bbq}]$ relates to the var limits that can be supplied or absorbed at all nodes in $\ccalD$, according to their physical limits or operational standards. 

The VVO problem \eqref{eq:vvo} is the basis for designing the optimal \textbf{volt-var control (VVC)} rules \cite{Kekatos2,IEEE1547}. Different from VVO that searches any feasible $\bbq$, the VVC problem seeks  a decision rule of $q_i \leftarrow g (v_i; \bbz_i)$ that is parameterized by $\bbz_i$, such that the $q_i$ decision at bus $i$ can be continuously updated based only on the localized voltage  $v_i$. This localized design greatly reduces computational and communication overhead, and thus is easy to implement in practice. We will discuss more details on the formulation of the VVC rule design problem  in Section \ref{sec:voltvar_ctrl}.  

\begin{figure*}[!t]
    \centering
    \includegraphics[width=\textwidth]{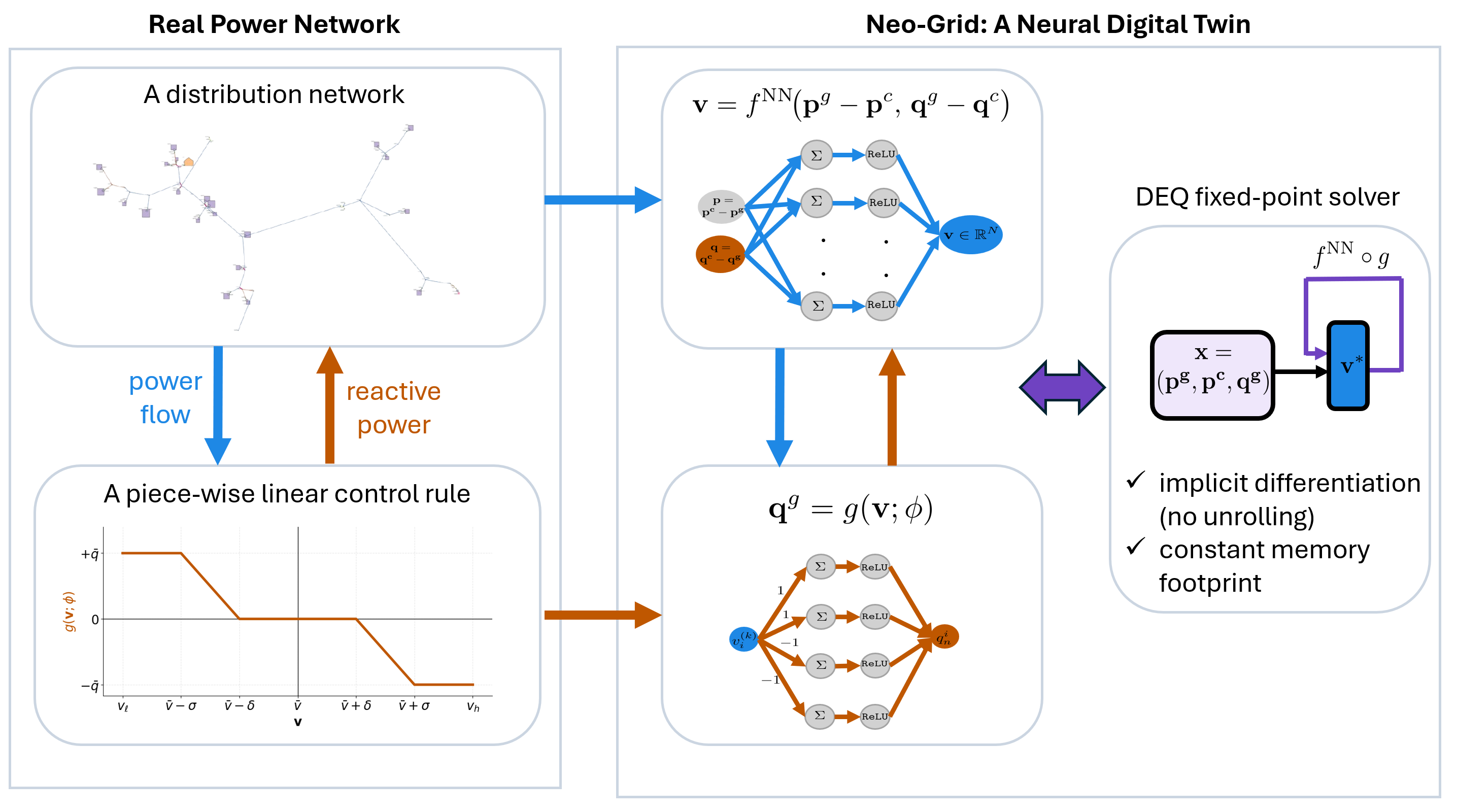} 
    \caption{Overview of \texttt{NEO-Grid} as a neural digital twin for optimizing the volt-var control (VVC) rules. 
    \textit{Left}: a distribution network controlled by piecewise-linear VVC rules at all nodes in $\ccalD$. 
    \textit{Center}: neural models for (top) the nonlinear PF equations of the voltage-power mapping 
    and (bottom) the VVC rules using a structured ReLU-based layers. 
    \textit{Right}: efficient ML solver to optimize the overall composited neural model 
    \((f^{\mathrm{NN}}\!\circ g)\) such that the fixed-point voltage \(\mathbf{v}^*\)  approaches 1.0 pu. 
    }
    \label{fig:neogrid_overview}
\end{figure*}

To solve either of these two problems, the key challenge lies in the non-linearity of the AC-PF model $f(\cdot)$. Thus,  this paper seeks to put forth a neural approximation approach to address this challenge using accurate yet computationally tractable PF models. The resultant \texttt{NEO-Grid} model serves as a surrogate of the underlying physical feeder, that can be conveniently incorporated into decision-making problems like VVO or VVC. \edit{Figure~\ref{fig:neogrid_overview} presents an overview of using \texttt{NEO-Grid} as a neural digital twin for VVC, based on a neural representation for the VVC rules too. This fully neural representation, as our key innovation, enables powerful ML techniques to efficiently optimize the VVC rules.}


\section{\texttt{NEO-Grid}: neural PF approximation}
\label{sec:learning_pf}
We provide a comprehensive overview of both linear approximation and our neural approximation for AC power flow (AC-PF) models, as used by distribution-level optimization and control tasks. 


\paragraph{Linear PF Approximation:}~Linearization is commonly used to simplify the AC-PF model. In particular, the LinDistFlow model approximates \eqref{eq:DF} by neglecting all loss-related terms and using $v_i^2 \approx 2v_i -1$ around $v_i \approx 1.0$~pu, leading to
\begin{align}
    \mathbf{v} &= \mathbf{R} \mathbf{p} + \mathbf{X} \mathbf{q} + \mathbf{1}v_0 \label{eq:ldf}
\end{align}
Here, $\mathbf{R}$ and $\mathbf{X}$ are the sensitivity matrices derived from the feeder topology and line impedance parameters. For the VVO problem in \eqref{eq:vvo}, it is often written as $\bbv = \bbX \bbq^g + \bar{\bbv}$, where $\bar{\bbv}$ denotes the nominal voltage profile without any var support  at $\bbq^g=\mathbf{0}$.  

To improve accuracy, data-driven linearization can be employed. Given training samples $\{[\bbp; \bbq], \bbv\}$, a least-squares (LS) regression model can be trained by minimizing the loss function:
\begin{align*}
    \| \mathbf{v} - f^\mathrm{LS} ([\bbp; \bbq]; \bbbeta) \|_2^2
\end{align*}
where $\bbbeta$ are the linear (affine) model coefficients. 
While linear approximations greatly simplify downstream optimization and control, their accuracy diminishes under substantial voltage drops, high losses, or diverse operating conditions.


\paragraph{NN-Based PF Approximation:} 
Neural approximations improve on linear models by learning a nonlinear power-voltage mapping from samples generated by various operating conditions. 
Traditionally, 
piecewise-linear (PWL) approximations 
need to use a manual partition of the input space into multiple linear regions \cite{PWL} and thus are often computationally intensive for the subsequent mixed-integer optimization. 
Neural networks (NNs) offer a scalable alternative: as universal function approximation, they can flexibly learn the nonlinear structure directly from samples without the need of explicit partitioning. The trainable ReLU activations in the hidden layers effectively encode the model’s nonlinear characteristics.
To train a neural approximation $f^\mathrm{NN}(\cdot)$, we perturb the nominal generation/demand profiles to create diverse samples of net active and reactive power $[\mathbf{p};~\mathbf{q}]$. The corresponding voltages $\mathbf{v}$ are obtained from accurate PF simulations. This ensures the resultant dataset covers both a wide range of operating scenarios, including the var compensation cases with possibly negative $\bbq$.

\edit{To design the \texttt{NEO-Grid} model, we adopt a one-hidden-layer NN architecture for simplicity; see the center-top panel in Figure~\ref{fig:neogrid_overview}.} Of course, deeper layers could be used to further improve the accuracy, but may incur increased complexity or data overfitting issues. The NN input is the net power injection vector of length $2N$, while the output is the voltage vector $\bbv$ of length $N$. The hidden layer consists of $K$ ReLU units and is parameterized by weight matrices $\mathbf{W}^{(1)}$, $\mathbf{W}^{(2)}$, and bias vector $\bbb$. Formally, the NN model becomes
\begin{align*}
    \mathbf{v} = f^\mathrm{NN}([\mathbf{p}; \mathbf{q}]; \bbtheta)= \mathbf{W}^{(2)} \cdot \mathrm{ReLU}\big(\mathbf{W}^{(1)} [\mathbf{p}; \mathbf{q}] + \bbb\big)
\end{align*}
with all trainable parameters in $\boldsymbol{\theta} = \{\mathbf{W}^{(1)}, \mathbf{W}^{(2)}, \boldsymbol{\beta}\}$.

\begin{remark}[General NN Output]
While our model directly predicts the voltage vector $\bbv$, other output choices such as the voltage deviations $|\bbv - \mathbf{1}|$ and the total deviation norm $\| \bbv - \mathbf{1} \|$ have been considered in the literature for specific control objectives. For example, \cite{Yize} has modeled the output of $|\bbv - \mathbf{1}|$ to allow for efficient voltage regulation.
\end{remark}

\begin{remark}[Compact \texttt{NEO-Grid} Design]
Although we focus on a fully connected NN here, it is advantageous to exploit the sparse feeder connectivity to build compact NN models with fewer parameters. Such topology-aware design, as advocated in \cite{Young-ho,Shaohui}, can improve the model generalizability and mitigate data overfitting. This direction is very important to explore in future.
\end{remark}


\subsection{VVO on \texttt{NEO-Grid} via mixed-integer programming}
\label{sec:voltvar_opt}



Integrating a neural approximation \( f^\mathrm{NN}(\cdot) \) into VVO has been advocated for scalable and data-driven voltage regulation; see e.g., \cite{Yize,ConstraintLearning,Young-ho}. By replacing the nonlinear AC-PF in \eqref{eq:vvo} with the PWL model based on  the trained NNs. This surrogate enables the embedding of PF constraints in mixed-integer linear (MIL) form that can be solved by off-the-shelf optimizers. Specifically, the VVO problem becomes
\begin{subequations}
\label{eq:vvo_nn}
\begin{align}
    \min_{\mathbf{q}^g} \quad & 
        \| \mathbf{v} - \mathbf{1} \|_2^2 \\
    \text{s.t.} \quad & 
        \mathbf{v} = f^\mathrm{NN}(\mathbf{p}^g - \mathbf{p}^c, \mathbf{q}^g - \mathbf{q}^c) \label{eq:vvo_nnf} \\
                      & 
        \underline{\mathbf{q}}^g \leq \mathbf{q}^g \leq \overline{\mathbf{q}}^g
\end{align}
\end{subequations}
%
%
Here, the controllable \( \mathbf{q}^g \) and resulting voltage \( \mathbf{v} \) are decision variables, with \( \mathbf{p}, \mathbf{q}^c \) fixed. 

The learned model \( f^\mathrm{NN}(\cdot) \) provides a differentiable power-voltage mapping. Constructed with ReLU-based neurons, \( f^\mathrm{NN}(\cdot) \) is continuous but piecewise-linear. Therefore, it introduces combinatorial structure through binary activation patterns, resulting in MIL constraints in \eqref{eq:vvo_nnf}; see more discussions on this reformulation in \cite{Young-ho}. With a quadratic objective,   the overall VVO problem is a mixed-integer quadratic program (MIQP). Moreover, \cite{Yize} has considered using a convex function design of \( f^\mathrm{NN}(\cdot) \) to further accelerate the VVO solution, but it is restricted to certain types of inequality constraints to remain a convex program. In general, the neural PWL surrogate enables embedding nonlinear PF physics into the optimization in a compact and computationally tractable manner, with a general and interoperable applicability.

\section{VVC on \texttt{NEO-Grid} via DEQ-based implicit learning}
\label{sec:voltvar_ctrl}


We now formally formulate the VVC rule design problem and present our solution by leveraging the \texttt{NEO-Grid} based surrogate nonlinear PF model. See the overall design in Figure~\ref{fig:neogrid_overview}. The VVC problem formulation follows from~\cite{Kekatos1,Kekatos2}. Unlike these existing methods, our solution employs the \textbf{deep-equilibrium model (DEQ)}~\cite{DEQ,MDEQ} to attain memory-efficient learning of the optimal VVC parameters. Computation efficiency is a key innovation of our work, which can eliminate the need of rolling out the recursive system dynamics and directly optimize the single-layer model by using DEQ.  

Following the IEEE 1547 standard~\cite{IEEE1547}, we represent the VVC rules for smart inverters as a \textbf{piecewise linear (PWL)} function.  It maps the local voltage measurement \( v_i \) at any inverter node \( i \in \ccalD \) to its reactive power output \( q^g_i \), \edit{as illustrated in the left--lower panel of Figure~\ref{fig:neogrid_overview}.}  This rule is parameterized by four interpretable quantities: nominal voltage setpoint \( \bar{v} \), deadband half-width $\delta$, ramp width \( \sigma \), and maximum reactive power magnitude \( \bar{q} \). The slope of the ramp regions is thus given by:
\begin{align}
    \alpha = \frac{\bar{q}}{\sigma - \delta} \label{eq:alpha}
\end{align}
where \( \sigma - \delta \) is the width of the ramp region outside the deadband.

The IEEE 1547 standard also imposes the following bounds on the parameters to ensure safe and interoperable operations:
\begin{subequations}
\label{eq:IEEE_standards}
\begin{align}
    0.95 &\leq \bar{v} \leq 1.05 \\
    0 &\leq \delta \leq 0.03 \\
    \delta + 0.02 &\leq \sigma \leq 0.18 \\
    0 &\leq \bar{q} \leq \hat{q}.
\end{align}
\end{subequations}

Using $\bbphi_i$ to represent these four parameters of node $i\in\ccalD$, we explicitly define the VVC control rule as:
\begin{align}
   \mathbf{q}^g = g(\mathbf{v}; \bbphi) \label{eq:vvc}
\end{align}
where \( g(\cdot; \bbphi) \) is the overall function by concatenating the PWL mapping per inverter node $i$ and  including all parameters in \( \bbphi := \{\bbphi_i\}_{i\in\ccalD} \). Note that the VVC rule operates in a closed-loop fashion: each inverter measures its local voltage \( v_i \) to compute \( q^g_i = g(v_i; \bbphi_i) \), and upon implementing inverter control to achieve the updated \( \bbq^g \), the feeder reaches a new operating point with voltage profile \( \bbv \) per the actual power flow model. This process iterates until the system converges to a steady-state voltage profile \( \bbv^* \). Note that this equilibrium can be  achieved very quickly with a negligible time period,  thanks to the fast responses of the power electronics based control in smart inverters~\cite{DrZhu}.

The \textbf{VVC rule design problem} — as formalized in~\cite{Kekatos1,Kekatos2} — is to determine the optimal parameter vector \( \bbphi \) such that the resulting \( \bbv^* \) approaches the flat profile at 1.0~pu as much as possible, given  by:
\begin{subequations}
\label{eq:rule_design}
\begin{align}
    \min_{\bbphi} \quad & \| \bbv^* - \mathbf{1} \|_2^2 \\
    \text{s.t.} \quad & \bbv^* = f\big( \bbp^g - \bbp^c, \bbq^g \leftarrow g(\bbv^*; \bbphi) - \bbq^c \big) \label{eq:vvc-eq} \\
                       & \text{IEEE standard bounds in~\eqref{eq:IEEE_standards}.} 
\end{align}
\end{subequations}
Here, \( f(\cdot) \) denotes the actual power flow mapping  and the constraint in \eqref{eq:vvc-eq} describes the equilibrium condition for the aforementioned VVC iterations based on the rule \( g(\cdot; \bbphi) \) in \eqref{eq:vvc}.

This formulation highlights the coupled nature between voltage and reactive power through \( g(\cdot) \) and \( f(\cdot) \).  Due to the presence of PWL mappings, it is common to introduce binary integer variables to produce the different linear segment of the VVC rules. This makes \eqref{eq:rule_design} a non-convex optimization problem with both continuous and combinatorial variables, in addition to the nonlinear PF model. As such, it becomes a mixed-integer nonlinear program (MINLP) and can pose computational challenges in large-scale settings. \edit{To address both the nonlinear PF and PWL rules, we put forth a fully neural representation by using both  \texttt{NEO-Grid}'s \(f^{\mathrm{NN}}\) and a structured ReLU realization of \(g(\cdot;\bbphi)\). This way, the closed-loop dynamics is fully characterized by a \emph{single implicit layer}, that is amenable to memory- and compute-efficient learning without unrolling; see the right panel of Figure~\ref{fig:neogrid_overview}. Before introducing the proposed solution, we remark on how to ensure the stability or convergence of the VVC rule-based closed-loop dynamics. } 

\begin{remark}[Stability of the Closed-Loop VVC Dynamics]
In addition to the bounds in \eqref{eq:IEEE_standards}, it is important to design the rule parameters to further ensure the stability and convergence of the closed-loop dynamics. Although a closed-form stability criterion is very challenging for the underlying nonlinear system, analyzing this under the linearized approximate system model is still meaningful for practical considerations~\cite{Kekatos1}. Note that the same analysis can be used by our neural surrogate based dynamics too, because both linear and neural models are approximates of the actual AC-PF. As detailed in~\cite{Kekatos1}, using the voltage-reactive power sensitivity matrix $\bbX$ in \eqref{eq:ldf}, one can use the sufficient spectral-norm condition, 
$ \| \bbD_{\bbalpha} \mathbf{X} \|_2 \le 1 - \epsilon$, to ensure stability. Here,  $\bbD_\bbalpha$ stands for the diagonal matrix formed by the ramp slope $\alpha_i$'s  at all nodes [cf.~\eqref{eq:alpha}]. For tractable optimization, 
one can use the polytopic relaxation 
\[
\alpha_i \le \frac{1-\epsilon}{\sum_{(i,j) \in \mathcal{E}} X_{ij}}, \ \forall i \in \mathcal{D}
\]
where the factional term uses all neighboring buses of node $i$ to include   
the coupling in the $i$-th row of matrix $\bbX$.  
This way, the closed-loop iterations are guaranteed to converge to the fixed-point equilibrium $\bbv^*$ under the linearized PF model. Since the latter is also an approximation, we can adopt the same constraints to promote stable closed-loop dynamics for our neural surrogate based iterations, as well.
\end{remark}


\subsection{Deep Equilibrium Learning for  VVC}

\edit{We present the detailed steps for representing the overall dynamics as a \emph{single implicit layer} and implementing the DEQ-based efficient learning, 
as outlined in Figure~\ref{fig:neogrid_overview}.}

\paragraph{NN-based Rule Representation:}
To achieve fully neural closed-loop dynamics, we first  parameterize the VVC rule $g(\cdot;\bbphi)$ as a structured ReLU-based NN with learnable breakpoints. This step is similar to~\cite{Kekatos1,Kekatos2}, and can preserve the PWL form of the standardized VVC rules while ensuring efficient differentiation by gradient-based algorithms. As illustrated in the center--bottom panel of \edit{Figure~\ref{fig:neogrid_overview}},  each VVC rule with local voltage input $v_i$ can be expressed as the sum of four ReLU-based PWL functions with respective breakpoints at
\((\bar{{v}}\pm\delta)\) and \((\bar{{v}}\pm\sigma)\). Each of them corresponds to either the saturation or ramp region and, together, they produce the overall VVC rule. This way, the VVC rule $g(\cdot;\bbphi)$  becomes equivalent to a two-layer NN with ReLU activation as shown in Figure \ref{fig:neogrid_overview}. 

\paragraph{Fixed-Point Condition.} 
Rather than unrolling the closed-loop dynamics to a recurrent sequence, we directly seek the steady-state voltage profile $\bbv^*$ as a fixed point:
\begin{align}
    \bbv^* &= \mathcal{F}_{\bbphi}(\bbv^*)
    \label{eq:fixed_point}\\
\textrm{with}
    ~\mathcal{F}_{\bbphi}(\bbv) &:= f^{\mathrm{NN}}\big(\mathbf{p}^c-\mathbf{p}^g,\; \mathbf{q}^c - g(\bbv;\bbphi)\big). 
\end{align}
Note that this is in a \textit{single implicit layer} form, by composing both NEO-Grid and NN-based rule into $\mathcal{F}_{\bbphi}$. Clearly, the choice of VVC rule parameter $\bbphi$ fully determines the fixed-point $\bbv^*$. 
To solve \eqref{eq:fixed_point}, we can use the Anderson acceleration based numerical solver~\cite{Anderson1965,WalkerNi2011} following \cite{DEQ}.

\paragraph{Memory-Efficient Implicit Differentiation.} 
We use the loss function given by
\begin{align}
    \mathcal{L}(\bbv^*;\bbphi) = \left\| \bbv^*(\bbphi) - \mathbf{1} \right\|_2^2,
    \label{eq:voltage_loss}
\end{align}
which is essentially the objective cost of voltage deviation in \eqref{eq:rule_design}. Note that this is an unsupervised learning design, as the loss function does not include any labeled target. It makes it very convenient to generate a large number of samples without solving the MINLP in \eqref{eq:rule_design}. 

To find the gradient wrt to $\bbphi$, we can define the adjoint vector ${\bblambda}$ by differentiating both sides of \eqref{eq:fixed_point}:
\begin{align}
    \Big(\bbI - \bbJ^{\mathcal{F}}\big|_{\bbv^*}\Big)^{\!\top}{\bblambda} = \frac{\partial \mathcal{L}}{\partial \bbv^*} = 2(\bbv^*-\mathbf{1})
    \label{eq:adjoint}
\end{align}
Using $\mathbf{\lambda}$ makes the gradient computation very efficiently, by using a vector-Jacobian product (VJP) as
\begin{align}
    \frac{\partial \mathcal{L}}{\partial \bbphi} = -{\bblambda}^\top \frac{\partial \mathcal{F}_{\bbphi}}{\partial \bbphi}\Big|_{\bbv^*}.
    \label{eq:param_grad}
\end{align}
Note that both \eqref{eq:adjoint} and \eqref{eq:param_grad} can be solved using the single layer form in \eqref{eq:fixed_point}, without rolling out the closed-loop dynamics. This makes the DEQ solver memory- and compute-efficient to directly optimize the equilibrium $\bbv^*(\bbphi)$. In practice, this implicit gradient in \eqref{eq:param_grad} is directly computed by autograd using the VJP without solving \eqref{eq:adjoint}.
Algorithm~\ref{alg:deq_training} summarizes the complete training procedure.

\begin{algorithm}[t]
\caption{DEQ-Based VVC Rule Design (mini-batch training)}
\label{alg:deq_training}
\KwIn{Scenarios $\{ \mathbf{x}_i \}_{i=1}^N$ with $\mathbf{x}_i=(\mathbf{p}^c_i,\mathbf{p}^g_i,\mathbf{q}^c_i)$; number of epochs $E$; batch size $B$; and learning rate $\eta>0$.}
\KwOut{Optimized VVC parameters $\bbphi$}

Initialize $\bbphi$ \\
\For{$e \gets 1$ \KwTo $E$}{
  Partition $\{1,\dots,N\}$ into mini-batches $\mathcal{B}_1,\ldots,\mathcal{B}_{\lceil N/B\rceil}$ \\
  \ForEach{mini-batch $\mathcal{B}$}{
    \ForEach{$i \in \mathcal{B}$}{
      Solve the fixed point $\bbv_i^* = \mathcal{F}_{\bbphi}(\bbv_i^*;\mathbf{x}_i)$ using Anderson acceleration
    }
    Evaluate the loss by $\mathcal{L}_{\mathcal{B}} \gets \frac{1}{|\mathcal{B}|}\sum_{i\in\mathcal{B}} \left\| \bbv_i^* - \mathbf{1} \right\|_2^2$ \quad [cf. Eq.~\eqref{eq:voltage_loss}] \\
    Compute $\nabla_{\bbphi}\mathcal{L}_{\mathcal{B}}$ via autograd based on Eqs.~\eqref{eq:adjoint}--\eqref{eq:param_grad} \\
    $\bbphi \gets \bbphi - \eta\, \nabla_{\bbphi}\mathcal{L}_{\mathcal{B}}$ \\
    Project $\bbphi$ onto the IEEE bounds in~\eqref{eq:IEEE_standards}
  }
}
\Return{$\bbphi$}
\end{algorithm}

\section{Numerical Results}
\label{sec:numerical_results}

\paragraph{Experimental Setup.} 
We evaluate all models and control strategies on the IEEE 33-bus radial distribution network \cite{IEEE33BW}, which consists of $N = 32$ load buses and one substation. This network is widely used in distribution system research due to its realistic topology and moderate complexity.

All simulations and experiments were conducted on a personal laptop (Dell Inspiron 16 Plus 7640) running Windows 11 Home (Build 26100), equipped with an Intel CPU (Intel64 Family 6 Model 170, $\sim$1.4~GHz), 32~GB of RAM, and an NVIDIA GeForce RTX 4060 Laptop GPU with 8~GB of VRAM.

Neural network models were implemented and trained using PyTorch. Data generation and power flow simulations were performed using the Pandapower library. All experiments and results are fully reproducible; the source code and datasets are publicly available on GitHub at:
\url{https://github.com/MFHChehade/Neo-Grid}.


\paragraph{Experimental Datasets.} 
We construct two distinct datasets to support the learning-based modeling and control evaluations. All data is generated using the \texttt{Pandapower} toolbox, using a fixed nominal operating point for the IEEE 33-bus network (see Section~3). At this nominal point, all active and reactive power values are interpreted as consumption. All quantities are expressed in per-unit (p.u.) on a 1~MVA base.

The \textbf{first dataset} is used to train the neural network power flow approximation model, as described in Section~\ref{sec:learning_pf}. Each data point consists of a 64-dimensional input vector, formed by concatenating the net active and reactive power at all $N = 32$ buses: $[\mathbf{p}; \mathbf{q}] \in \mathbb{R}^{64}$. The output is the corresponding bus voltage vector $\mathbf{v} \in \mathbb{R}^{32}$, computed via an AC power flow simulation. Thus, each row of the dataset contains 96 columns (64 inputs + 32 voltage outputs). 

This dataset contains 20{,}000 samples, with an 80/20 train-test split. The net power demands are generated by randomly perturbing the nominal active and reactive power values within $\pm 10\%$. Importantly, to ensure that the learned voltage model generalizes to scenarios with reactive power compensation (i.e., $\mathbf{q^g} < 0$), we inject an additional layer of perturbation to the reactive power entries: each bus's net $\mathbf{q}$ is randomly perturbed within the range $[-0.8, +0.2]$~p.u. This enables the NN model to learn accurate voltage responses in settings where reactive generation may be present. This additional perturbation applies to both training and test splits of the dataset. Once trained and validated, the neural network no longer uses this dataset in downstream control tasks.

The \textbf{second dataset} is used exclusively for evaluating and training VVO and VVC schemes. It contains 100 samples, generated with the same $\pm 10\%$ perturbation to nominal active and reactive power demands, but without any artificial correction to the reactive power. As such, the reactive power profile remains consistent with realistic consumption scenarios, and will instead be adjusted by the VVO and VVC algorithms. As before, each sample yields a total of 96 columns via power flow simulation.

The dataset is again split 80/20. The first 80 samples are used to train the VVC control rule parameters (see Section~\ref{sec:voltvar_ctrl}), while the final 20 samples are used to evaluate both VVO and VVC performance under unseen loading conditions. The VVO method, being optimization-based, does not require a training phase.

\begin{table}[t]
    \centering
    \renewcommand{\arraystretch}{1.2}
    \setlength{\tabcolsep}{8pt}
    \caption{\textbf{Test set voltage prediction error (MSE) for different power flow approximation models.} The neural network significantly outperforms both LinDistFlow and linear regression, validating the importance of nonlinear modeling in accurately capturing power flow behavior.}
    \begin{tabular}{lc}
        \toprule
        \textbf{Model} & \textbf{Voltage MSE (Test Set)} \\
        \midrule
        LinDistFlow & $1.95 \times 10^{-4}$ \\
        Linear Regression (LS) & $3.16 \times 10^{-5}$ \\
        \textbf{NEO-Grid (NN)} & $\mathbf{1.81 \times 10^{-6}}$ \\
        \bottomrule
    \end{tabular}
    \label{tab:pf_mse}
\end{table}

\subsection{Power Flow Approximation Performance.} 
To evaluate the accuracy of learning-based voltage modeling, we train a neural network to approximate the mapping from bus-level power consumption to voltage magnitudes. Specifically, the model takes as input the net real and reactive power demands at all $N = 32$ buses, forming a 64-dimensional input vector $[\mathbf{p}; \mathbf{q}]$, and outputs the 32-dimensional voltage vector $\mathbf{v}$. The training is performed on the first dataset described previously, which covers a wide range of operating conditions via $\pm10\%$ perturbations and reactive compensation ranging from $-0.8$ to $+0.2$ p.u.

\paragraph{Baseline Models.} 
We benchmark the proposed neural network model against two standard alternatives. First, we use the \textbf{LinDistFlow} model, a physics-inspired linear approximation derived from first principles, which requires no data-driven  \cite{IEEE33BW}. Second, we implement a \textbf{linear regression} model, equivalent to a neural network with no hidden layer. This serves as a data-driven linear baseline trained on the same perturbed dataset, yielding a single affine mapping from power injections to voltages. These comparisons help isolate the benefits of nonlinearity in neural modeling.

\paragraph{Neural Network Setup.}
The neural network used for voltage prediction consists of one hidden layer with $K = 64$ ReLU units. The model is trained using PyTorch and optimized using the Adam optimizer with a learning rate of $10^{-4}$, over 50{,}000 epochs with batch size 1. Input and output data are scaled to $[0,1]$ using a min-max normalization fitted to the data. 

\paragraph{Results.}
We evaluate each model on the 20\% test split (4{,}000 samples), computing the mean squared error (MSE) between predicted and true voltages over all nodes. As shown in Table~\ref{tab:pf_mse}, the neural network achieves an MSE of $1.81 \times 10^{-6}$, which is an order of magnitude better than the linear regression model ($3.16 \times 10^{-5}$) and over 100x better than the LinDistFlow approximation ($1.95 \times 10^{-4}$).

These results confirm that learning a nonlinear mapping using a hidden layer with ReLU activation yields a much more expressive and accurate model of the power flow dynamics. While the LinDistFlow model is analytically convenient and the linear regression model adapts to data, both fail to capture the nonlinear coupling between nodal powers and voltages that arise due to network topology and line impedances. The neural network, in contrast, learns to accurately account for these effects, making it a more suitable surrogate for downstream optimization and control.
\begin{table*}[t]
    \centering
    \renewcommand{\arraystretch}{1.2}
    \setlength{\tabcolsep}{12pt}
    \caption{
        \textbf{Voltage deviation statistics for different volt-var optimization models across 20 test scenarios.}
        The first column shows the average absolute deviation from 1.0~p.u., while the remaining columns report the percentage of voltage entries exceeding 1\%, 3\%, and 5\% deviation thresholds, respectively. All optimization-based approaches significantly outperform the no-correction baseline. The neural network model (NEO-Grid) achieves the best performance across all metrics, including the lowest average error and the smallest rates of large violations.
    }
    \begin{tabular}{lcccc}
        \toprule
        \textbf{Model} & \textbf{Avg Deviation} & \textbf{$>1\%$} & \textbf{$>3\%$} & \textbf{$>5\%$} \\
        \midrule
        \textbf{NEO-Grid (NN)} & \textbf{0.94\%} & \textbf{28.13\%} & \textbf{5.47\%} & \textbf{0.00\%} \\
        Linear Regression (LS) & 0.97\% & 31.25\% & 10.78\% & 0.00\% \\
        LinDistFlow & 1.04\% & 30.47\% & 11.72\% & 0.00\% \\
        No Correction ($\mathbf{q}^g = \mathbf{0}$) & 5.32\% & 84.38\% & 71.88\% & 65.31\% \\
        \bottomrule
    \end{tabular}
    \label{tab:vvo_deviation}
\end{table*}

\subsection{VVO Test Results.} 

\paragraph{Optimization Setup.}
We solve the volt-var optimization problem formulated in \eqref{eq:vvo} for the 20 testing samples in the second dataset. The optimization is implemented using the \texttt{Pyomo} framework \cite{Pyomo}, with the neural network power flow surrogate embedded via the \texttt{OMLT} toolkit \cite{OMLT}. We enforce inverter reactive power constraints by bounding the decision variables $\mathbf{q}^g$ within $[-0.6, 0.1]$~p.u. at designated DER nodes. 

To improve tractability, we approximate the quadratic objective $\| \mathbf{v} - \mathbf{1} \|_2^2$ using a linear surrogate based on the $\ell_1$ norm: $\| \mathbf{v} - \mathbf{1} \|^2_1$. This relaxation enables solving the neural VVO problem as a mixed-integer linear program (MILP) rather than a computationally expensive MIQP. The same $\ell_1$ formulation is used across all models to ensure comparability. All problems are solved using the open-source \texttt{CBC} solver \cite{CBC}, which supports mixed-integer linear optimization.

\paragraph{Results and Analysis.}
We evaluate each method by computing voltage deviation from the nominal 1.0~p.u. profile across all buses and test scenarios. We report: (i) the average absolute deviation across all bus-voltage values; and (ii) the percentage of voltage entries with deviation exceeding 1\%, 3\%, and 5\%. These statistics are computed over $20 \times 32 = 640$ voltage entries. Table~\ref{tab:vvo_deviation} summarizes the results.

The results clearly show that all three optimization-based methods significantly improve voltage regulation compared to the no-correction baseline, which yields widespread violations. Among the methods, the NEO-Grid neural network model consistently achieves the best performance. It not only delivers the smallest average absolute voltage deviation (0.94\%) but also substantially reduces the incidence of large voltage violations: only 5.47\% of buses exceed a 3\% deviation, compared to over 10\% for linear regression and more than 11\% for LinDistFlow. No model exceeds the 5\% threshold, confirming the effectiveness of all methods in enforcing standard voltage limits. These findings emphasize the value of nonlinear modeling via neural networks in distribution system control, especially in minimizing extreme deviations that could compromise system reliability.


\begin{table*}[t]
    \centering
    \renewcommand{\arraystretch}{1.2}
    \setlength{\tabcolsep}{12pt}
    \caption{
        \textbf{Voltage regulation performance of different VVC models under 20 test scenarios.}
        The first column reports the average absolute deviation from the nominal 1.0~p.u. voltage profile. The next two columns show the percentage of bus voltages with less than 5\% and 7\% deviation, respectively. The DEQ-based NEO-Grid model achieves the lowest average deviation and eliminates all violations at and beyond 7\%, highlighting its superior ability to learn effective volt-var control policies. All optimization-based strategies improve substantially over the initial and no-correction baselines.
    }
    \begin{tabular}{lccc}
        \toprule
        \textbf{Model} & \textbf{Avg Deviation} & \textbf{ $>$ 5\%} & \textbf{ $>$ 7\%} \\
        \midrule
        \textbf{NEO-Grid (NN)} & \textbf{3.63\%} & \textbf{38.59\%} & \textbf{0.00\%} \\
        LinDistFlow & 4.66\% & 53.12\% & 28.12\% \\
        Linear Regression (LS) & 4.55\% & 50.00\% & 26.56\% \\
        Initial Weights (No Optimization) & 4.76\% & 53.12\% & 18.75\% \\
        No Correction ($\mathbf{q}^g = \mathbf{0}$) & 5.32\% & 65.31\% & 43.59\% \\
        \bottomrule
    \end{tabular}
    \label{tab:vvc_deviation}
\end{table*}

\subsection{Optimal VVC Test Results.} 

\paragraph{DEQ-Based VVC Implementation.}
To implement volt-var control within the \texttt{NEO-Grid} framework, we adopt the Deep Equilibrium Model (DEQ) architecture as described in~\cite{Zico_blog}. We design a minimalist DEQ structure with a single equilibrium layer responsible for modeling the closed-loop voltage dynamics induced by the control rules. No additional feedforward layers are used. The DEQ layer maps an initial voltage estimate $\mathbf{v}^{(0)}$ and an exogenous input $\mathbf{x} = (\mathbf{p}^c, \mathbf{p}^g, \mathbf{q}^c)$ to a fixed-point solution $\mathbf{v}^*$, which satisfies the condition:
\[
     \bbv^* = \mathcal{F}_{\bbphi}(\bbv^*, \mathbf{p}, \mathbf{q}^c)
\]

\paragraph{DEQ Training Details.}
Training is conducted on the first 80 samples of the second dataset, using batch training over mini-batches of size 16. The total number of training epochs is 500, with a learning rate of $10^{-3}$. The training process follows the batch-style DEQ optimization scheme introduced in~\cite{Zico_blog}, where equilibrium is approximated per batch using Anderson acceleration. To enforce the feasibility of learned control rules with respect to the IEEE 1547 standard (see Eq.~\eqref{eq:IEEE_standards}), we apply \textbf{projected gradient descent} at each training step. This ensures that parameter updates are immediately projected back onto the feasible region, a process that is readily incorporated into the PyTorch optimization pipeline.

\paragraph{Results and Analysis.}
Table~\ref{tab:vvc_deviation} summarizes the performance of various volt-var control strategies across 20 test scenarios. The DEQ-trained NEO-Grid controller yields the \textbf{lowest average deviation of 3.63\%}—outperforming all alternatives—and successfully eliminates all voltage entries with deviations at and beyond 7\%. This is in stark contrast to the other optimization-based models (LinDistFlow and Linear Regression), which still experience notable large deviations beyond 5\% and 7\%. Notably, while LinDistFlow and LS reduce overall error relative to the initial or uncontrolled profiles, they fail to match the robustness and consistency achieved by the learned DEQ controller. The no-correction baseline ($\mathbf{q}^g = \mathbf{0}$) performs the worst across all metrics, with more than 65\% of voltage entries deviating by over 5\%, confirming the necessity of reactive control. Overall, these results underscore the ability of equilibrium-based neural control design to deliver high-accuracy regulation under diverse load and generation conditions.

\paragraph{VVO versus VVC Performance.}
It is worth noting that the VVO formulation achieves consistently better voltage regulation than the optimal VVC methods across all metrics. This is expected: in VVO, the inverter reactive power injections $\mathbf{q}^g$ are treated as free decision variables constrained only by physical bounds, allowing the optimization to select values that minimize deviation directly. In contrast, VVC policies are restricted to follow a learned piecewise-linear control rule with a fixed structure—parameterized by $(\bar{v}, \sigma, \delta, \bar{q})$—that maps voltage measurements to reactive power responses. As such, even under optimal parameter tuning, the VVC framework cannot fully match the flexibility and performance of the unconstrained VVO problem. This highlights the inherent trade-off between policy interpretability and closed-loop control fidelity.

\section{Conclusions}
\label{sec:conclusion}
This paper presented \texttt{NEO-Grid}, a unified learning-based framework for volt-var optimization and control in distribution systems. By integrating nonlinear neural surrogates for power flow and equilibrium-based control via deep equilibrium models (DEQs), our approach captures closed-loop voltage dynamics while remaining tractable for both optimization and learning. Experiments on the IEEE 33-bus network show that NEO-Grid consistently outperforms traditional linear and heuristic methods in voltage deviation and constraint satisfaction. Future work includes extending the framework to multi-phase systems, stochastic inputs, and real-time deployment.

\balance


\section*{References}
\vspace{-10pt}
\bibliography{style/IEEEabbrv,ref}
\bibliographystyle{IEEEtran}


\end{document}